\documentclass{aa}  

\usepackage{graphicx}
\usepackage{txfonts}

\begin{document}

\title{Beyond self-similarity: Reconciling X-ray scaling relations \\
in galaxy clusters and groups}


%

   \author{S. Ettori\inst{1,2}\fnmsep\thanks{Corresponding author: stefano.ettori@inaf.it}
        }

   \institute{INAF, Osservatorio di Astrofisica e Scienza dello Spazio, via Piero Gobetti 93/3, 40129 Bologna, Italy \and
INFN, Sezione di Bologna, viale Berti Pichat 6/2, 40127 Bologna, Italy}

   \date{Accepted: 09/11/2026}

 
  \abstract 
   {Scaling relations hold among observed quantities that describe the thermodynamic properties of the gas in galaxy clusters and groups. However, observed data show systematic departures from the self-similar model's baseline predictions, particularly in lower-mass systems.}
   {I show that the observed departures from self-similar predictions can be efficiently described by two physical quantities modeled with power laws:
   the gas mass fraction ($f_g \sim T^{f_1} E_z^{f_z}$) and the temperature variation ($f_T \sim T^{t_1} E_z^{t_z}$). 
   These phenomenological parameters allow us to compact all X-ray scaling relations into a single generalized form.} 
   {Using a wide range of published X-ray scaling relations, this study conducts a Markov chain Monte Carlo--based meta-analysis to constrain the temperature- and redshift-dependence of the meta-parameters $f_g$ and $f_T$ and calibrate the model.}

   {These calibrations indicate that, while the gas mass fraction ($f_g$) does not show significant evolution with cosmic time ($f_z = -0.11 \pm 0.03$), it decreases significantly with decreasing halo mass ($f_1 = 0.50 \pm 0.01$). On the other hand, the temperature variation ($f_T$) shows a mild positive increase with both mass and redshift ($t_1 = 0.15 \pm 0.01$; $t_z = 0.11 \pm 0.07$). 
   A systematic uncertainty of $\sim 0.05$ can be assigned to $f_1$ and $t_1$, reflecting residual differences in instrument calibration, whereas statistical errors remain dominant for $f_z$ and $t_z$. 
   Overall, modeling the departures from the self-similar model with $\{f_g, f_T\}$ drastically improves predictive accuracy, reducing the number of scaling relations in $>3\sigma$ ($>5 \sigma$) tension from 49 (36) percent under the self-similar scenario to just 8 (3) percent (three and one out of 39, respectively) that might be identified for their peculiarity.
   By imposing a minimal relative statistical error of 3 (5) percent on the published relations, I reduce the number of relations in tension at $>3\sigma$ to two (none).
   Moreover, modeling through this generalized framework allows for an extended discussion of the expected slopes and redshift evolution for several X-ray scaling laws, including the new quantity
   $Y_{LGT0} = L^{-1} M_g^2 T^{a_L}$, a proxy for the cluster's volume that is independent of $f_g$ and $f_T$ by construction, and is predicted to scale directly with mass with a well-defined redshift evolution ($M \sim Y_{LGT0} f_g^0 f_T^0 E_z^2$), as validated against hydrodynamical simulations and observational constraints.
   }
   {}


   \maketitle
   \nolinenumbers

\defcitealias{ettori23}{E23}

\section{Introduction}

In the self-similar model of galaxy cluster formation and evolution, the thermodynamic properties of the X-ray-emitting plasma are predicted to depend solely on the halo mass ($M$) and redshift ($z$).
The model is based on a gravity-dominated scenario in which clusters are essentially scaled versions of one another. It predicts specific power-law scaling relations between integrated quantities such as mass, temperature, gas mass, and luminosity. 
Any redshift evolution is governed by the factor $E_z = [\Omega_{\Lambda} + \Omega_m(1+z)^3]^{0.5}$, which represents the evolution of the Hubble constant with cosmological parameters $\Omega_m$ and $\Omega_{\Lambda}$, indicating the mass and dark energy densities with respect to the critical value $\rho_{\rm c}$, respectively.
Under this model, characteristic thermodynamic quantities within an overdensity, $\Delta = M_{\Delta} / (4/3 \pi R_{\Delta}^3 \rho_{\rm c})$, scale according to the mass, $M_{\Delta}$, and redshift, $z$, of the halo
(see e.g. the reviews in \citealt{voit05}, \citealt{boehringer12}, and \citealt{ettori23}, the third of which is hereafter referred to as E23).

While the self-similar model provides a baseline, observed data show systematic departures, particularly in lower-mass systems \citep[$\sim 10^{13} - 10^{14} M_\odot$; see recent reviews in][]{eckert21,lovisari21,opp21}. 
These deviations are attributed to non-gravitational processes such as feedback from active galactic nuclei (AGNs) and star formation, which impact the distribution of the intra-cluster medium (ICM). 
To account for these departures, E23 present a semi-analytic model, $i(cm)z$, which introduces temperature-dependent and redshift-evolving parameters (namely, $f_g$ and $f_T$) to modify the ideal self-similar relations.
These parameters, $f_g$ and $f_T$ (see Sect.~2.1 and 2.2 in E23), are introduced as phenomenological but physically motivated factors that encode deviations from the self-similar model, thereby accommodating observed departures and generalizing the X-ray scaling relations in galaxy clusters.

The (meta-)parameters that I adopt and propagate through the scaling relations are:
\begin{itemize}
\item $f_g$ (gas mass fraction): This parameter represents the gas mass fraction within a fixed overdensity (hereafter $\Delta = 500$ times the critical density of the Universe at the cluster's redshift), defined as the ratio of the gas mass to the total mass ($M_{\rm g}/M$). To represent deviations in the gas distribution from the “closed-box” expectation \citep[see e.g.][]{angelinelli22}, it is expressed as a power-law dependence on temperature ($T$) and the evolution factor ($E_z$) in the form
\begin{equation}
f_g = f_0 T^{f_1} E_z^{f_z}.
\label{eq:fg}
\end{equation}

\item $f_T$ (temperature variation): This parameter describes the variation between the value of the gas temperature that should be considered and the one adopted in a scaling relation; it is described extensively in Sect.~\ref{sect:ft} and is modeled, as the gas mass fraction is, with a power-law dependence:
\begin{equation}
f_T = t_0 T^{t_1} E_z^{t_z}.
\label{eq:ft}
\end{equation}
\end{itemize}
It is worth noting that the temperature, $T$, in both equations refers to a ``measured'' quantity, such as the spectroscopic estimate $T_{\rm sp}$, and is here adopted as the reference quantity due to its intensive nature, which does not depend on any cosmological parameter.

Assuming that published scaling relations reflect the intrinsic properties of the underlying galaxy group and cluster populations, and thus represent different projections of the same physical quantities, I can use their best-fit values to calibrate my model and its parameters to account for departures from self-similarity.
E23 did this against 17 observed X-ray scaling laws from various cluster samples (including Planck ESZ, eFEDS, and SZ-selected objects), and determined the following constraints for these parameters:
$f_1 =  0.403 \pm 0.009; f_z = -0.004 \pm 0.023; \, 
t_1 = 0.144 \pm 0.017; t_z = 0.349 \pm 0.059$.

These calibrations indicate that while the gas mass fraction, $f_g$, does not show significant evolution with cosmic time ($f_z \approx 0$), it decreases significantly as halo mass decreases. 
Conversely, $f_T$ shows a mild positive evolution with redshift.
These parameters enable the model to accurately predict the slopes and redshift evolution of various thermodynamic quantities (such as luminosity, mass, and pressure).

In the present work, I revise this calibration according to the largest set of published scaling laws and adopt a more robust MCMC analysis. 
To refine the constraints on the four (meta-)parameters describing the departures from self-similarity, namely $\{f_1, f_z; t_1, t_z\}$ (see Eqs.~\ref{eq:fg} and \ref{eq:ft}, respectively), I use summary statistics collected from published work (see App.~\ref{app:pub}) on the best-fit values of the slopes, $B$, and evolution, $C$, for a large variety of scaling relations represented by the form
\begin{equation}
\log(Y) = A +B \log(X) +C \log(E_z).
\label{eq:scalaw}
\end{equation}

The normalization, $A$, is not considered in the present analysis, despite bringing a lot of physical information, as I highlight in E23 \citep[see, for more details][]{ettori15,ettori20}, but also several systematics depending, for example, on the selection function, calibration, and statistical analysis.
The slope, $B$, and evolution, $C$, are less sensitive to these systematics, but they are not fully immune: flux-limited selection and Malmquist bias can tilt the observed slope, and residual inter-telescope calibration offsets in temperature translate directly into slope offsets at the few-percent level \citep[][]{schellenberger+15,migkas20}.

I start with a discussion on the physical meaning of $f_T$, followed by an extended description of the ``generalized form'' (GF) of the X-ray scaling relations, which can accommodate both the self-similar scenario and its departures from self-similarity, including their redshift evolution. I conclude with the new calibration of the meta-parameters $\{f_g, f_T\}$.

\section{The physical meaning of $f_T$}
\label{sect:ft}

While the origin, role, and impact of the gas mass fraction, $f_g$, are self-explanatory and reflect the relative impact of the gas mass at different mass scales, the meaning of $f_T$ is more subtle.
It is justified by the evidence that the gas temperature varies radially and is expected to differ from any global value representative of, for example, a typical halo mass.
In other words, $f_T$ captures deviations from the basic assumption of isothermality arising from the radial temperature structure of the ICM.

As described in E23, if the temperature used in the scaling laws, as is usually done in this kind of study, is the spectroscopic measurement $T_{\rm sp}$, then $f_T = T(R_{500}) / T_{\rm sp}$, where $T(R_{500})$ is the 3D gas temperature at $R_{500}$ or, for an isothermal gas, $T_{500}$.
The latter, for a singular isothermal sphere of gas in hydrostatic equilibrium within a typical overdensity of total mass, can be expressed as \citep[see also, e.g.][]{voit05}
\begin{equation}
    T_{500} = \frac{G M_{500} \mu m_p}{2 R_{500}} = 8.949 \times \left( \frac{M_{500}}{10^{15} M_{\odot}} \right)^{2/3} \times E_z^{2/3}  \; {\rm keV}.
\end{equation}

\begin{figure}[ht]
   \centering
   \includegraphics[width=0.5\textwidth,trim=28mm 16mm 25mm 80mm, clip]{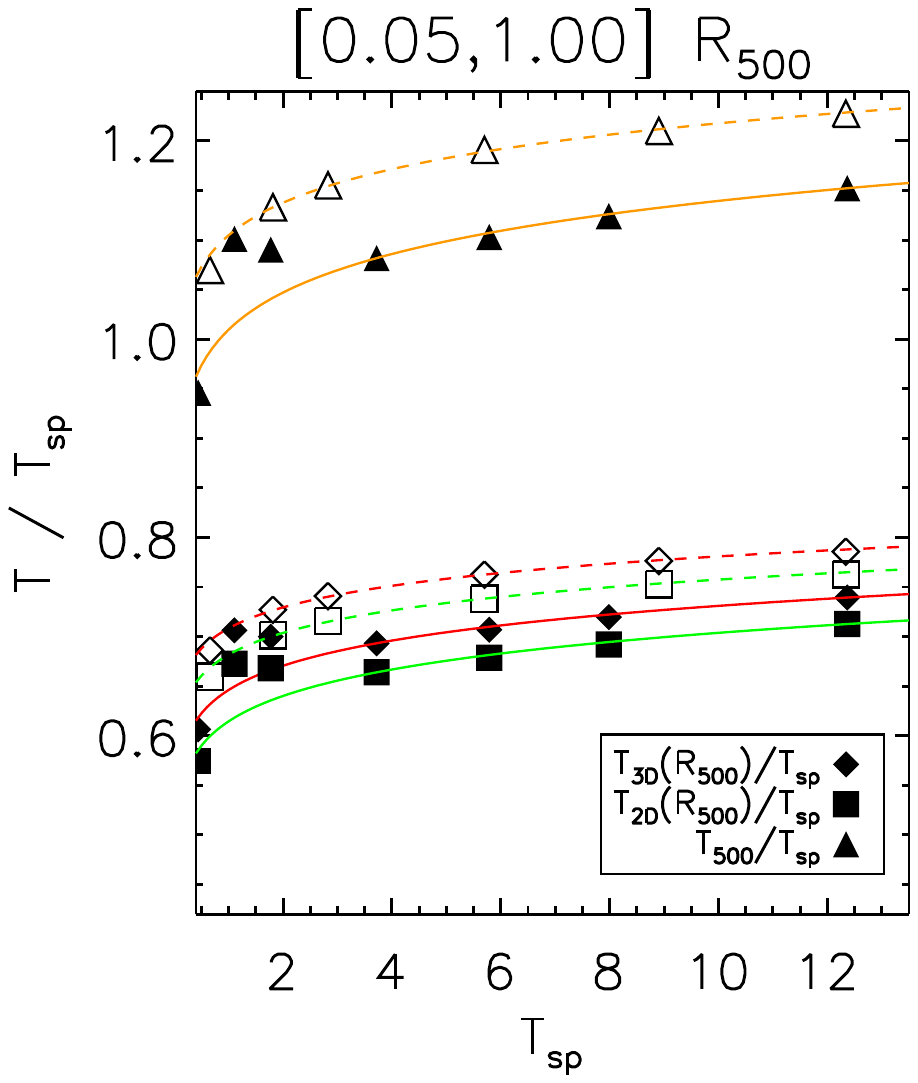} 
      \caption{For a set of $\{M_{500}, z\}$ values in the range $0.1-20 \times 10^{14} M_{\odot}$ and 0.01$-$1.4, respectively, I represent here (filled symbols at $z=0.01$; open symbols at $z=1.4$): $T_{3D}(R_{500})/T_{\rm sp}$ (diamonds); $T_{2D}(R_{500})/T_{\rm sp}$ (squares); $T_{500}/T_{\rm sp}$ (triangles).  The lines show the best fit with a power law discussed in the text.
      } \label{fig:ft}
   \end{figure}

Other ``global'' measurements refer to the weighted mean of the temperature values integrated along the line of sight:
\begin{equation}
    T_{wm} = \frac{\sum_i w_i T_i dV}{\sum_i w_i dV}.
    \label{eq:tmw}
\end{equation}
If a three-dimensional radial profile of the gas temperature is available, under the assumption of spherical symmetry, $T_i = T_{3D}(r)$. 
In the temperature regime considered in the present work and for estimates obtained with the current (CCD-based) X-ray telescopes, a reasonable weight to reproduce the spectroscopic temperature, $T_{\rm sp}$, is $w_i = n_{\rm gas}^2 T_{3D}(r)^{-0.75}$, which defines the spectroscopic-like value, $T_{\rm SL}$, described in \cite{mazzotta+04} 
that has been shown to be an effective representation of $T>3$ keV plasma, and slightly biased high (up to $\sim0.5$ keV) at lower temperatures 
\citep[see e.g. Fig. 7 in][]{vikhlinin06t}.
Similarly, the gas-mass weighted value, $T_{\rm mg}$, was obtained by defining $w_i = n_{\rm gas}$.

Moreover, the projected temperature profile will be different from $T_{3D}(r)$.
For instance, \cite{vik06} show that a simple approximated relation stands between the peak, spectroscopic average, and gas-mass-weighted temperatures (all measured in the radial range 70 kpc-$R_{500}$):  
$T_{\rm peak} : T_{\rm sp} : T_{\rm mg} = 1.21 : 1.11 : 1$.

To investigate the level of deviations in these measurements, I used
the $i(cm)z$ model, described in E23 and based on a universal pressure profile in equilibrium in a halo with a dark matter distribution following a concentration--mass-redshift relation.
Using it for a set of halo masses and redshifts, I computed the expected 3D and projected (2D) temperature profiles, and the corresponding $T_{500}$ and $T_{\rm sp}$.
The latter was calculated by summing the thermal emission described by an {\tt apec} model in {\tt XSPEC} \citep[vers. 12.15.1,][]{xspec} associated with ten components along the line of sight that are characterized by the corresponding $T_{3D}(r)$ and emission measure.
The overall spectrum simulated for an {\it XMM-Newton}/pn exposure was then fit with a single-temperature model with a fixed metallicity of 0.3 times the solar abundance, as in \cite{aspl09}.
 
In Fig.~\ref{fig:ft}, I plot the main outcomes of this analysis.
By integrating along the line of sight the radial values between $0.05-1 \, R_{500}$, I obtain $T_{3D}(r)/T_{\rm sp} \propto T_{\rm sp}^{0.05} E_z^{0.04}$, with $T_{500}/T_{\rm sp}$ behaving very similarly, and measure median values (min, max) of 
$T_{\rm peak} : T_{\rm sp} : T_{\rm mg} = 1.60 \, (1.41, 2.01) : 1.17 \, (1.14, 1.29) : 1$ at low redshift, 
and $1.30 \, (1.23, 1.49) : 1.14 \, (1.11, 1.22) : 1$ at $z=1.4$.
To reproduce most of the data that have spectral information available to a fraction of $R_{500}$, the same calculations done over the interval $0-0.5 \, R_{500}$ provide the following results: $T_{3D}(r)/T_{\rm sp}$ (and $T_{500}/T_{\rm sp}$) $\propto T_{\rm sp}^{0.07} E_z^{0.06}$; median values (min, max) of $T_{\rm peak} : T_{\rm sp} : T_{\rm mg} = 1.50 \, (1.38, 1.97) : 1.19 \, (1.15, 1.32) : 1$ at low redshift, and $1.31 \, (1.24, 1.44) : 1.14 \, (1.10, 1.23) : 1$ at $z=1.4$.
These results demonstrate that simple analytic models are also able to match the observed spectral properties of the ICM, providing at the same time strong evidence that any measurement of the ``global'' temperature is not unique, depending on the definition adopted (e.g., from the virial theorem, mass-weighted, spectroscopic one, and on the radial range covered).

It is worth noting that $f_T$ serves as a phenomenological envelope that captures, at least, three distinct physical effects. Two are the dominant contributions modeled by $i(cm)z$ and represented in Fig.~\ref{fig:ft}: (i) the radial temperature gradient due to the decline of the 3D temperature profile with radius, so $T_{\rm sp}$ underestimates $T_{500}$ systematically; (ii) projection effects, causing the mixing of the hotter and cooler gas along the line of sight, as quantified in the 2D profiles.
A third effect arises from the ICM thermalization state: the degree of thermalization depends on the system's dynamical age \citep[see, e.g.,][]{chen19}. At fixed mass and redshift, recently assembled clusters are cooler because they have had less time to virialize. Given cosmic formation history, more massive systems at fixed redshift have also formed more recently on average, introducing a mass- (and T-)dependent offset contributing to $t_1$.  
Because $f_T$ absorbs all three contributions, its calibrated parameters ($t_1$, $t_z$) reflect a combination of structural, dynamical, and observational effects not fully separable with the present meta-analysis.
Therefore, the use of the spectral temperature 
in a scaling law introduces biases that affect the measured normalization, slope, and redshift evolution.
I address these biases by modeling them through the term $f_T$, focusing in particular on their effects on the slope (parametrized by $t_1$) and evolution ($t_z$).

\section{The generalized form of the scaling relations}
\label{sect:gen}

In \cite{ettori15} (as an extension of the GF first presented in \citealt{ettori13}), I discussed a contracted form of the scaling relations where observables such as X-ray luminosity, $L$, gas temperature, $T$, and gas mass, $M_g$, are related to the halo mass, $M$, and redshift (through $E_z$) in their self-similar behavior, corrected by a model of the departure from self-similarity itself.
\begin{figure}[thb]
   \centering
   \includegraphics[width=0.5\textwidth,trim=12mm 10mm 28mm 85mm, clip]{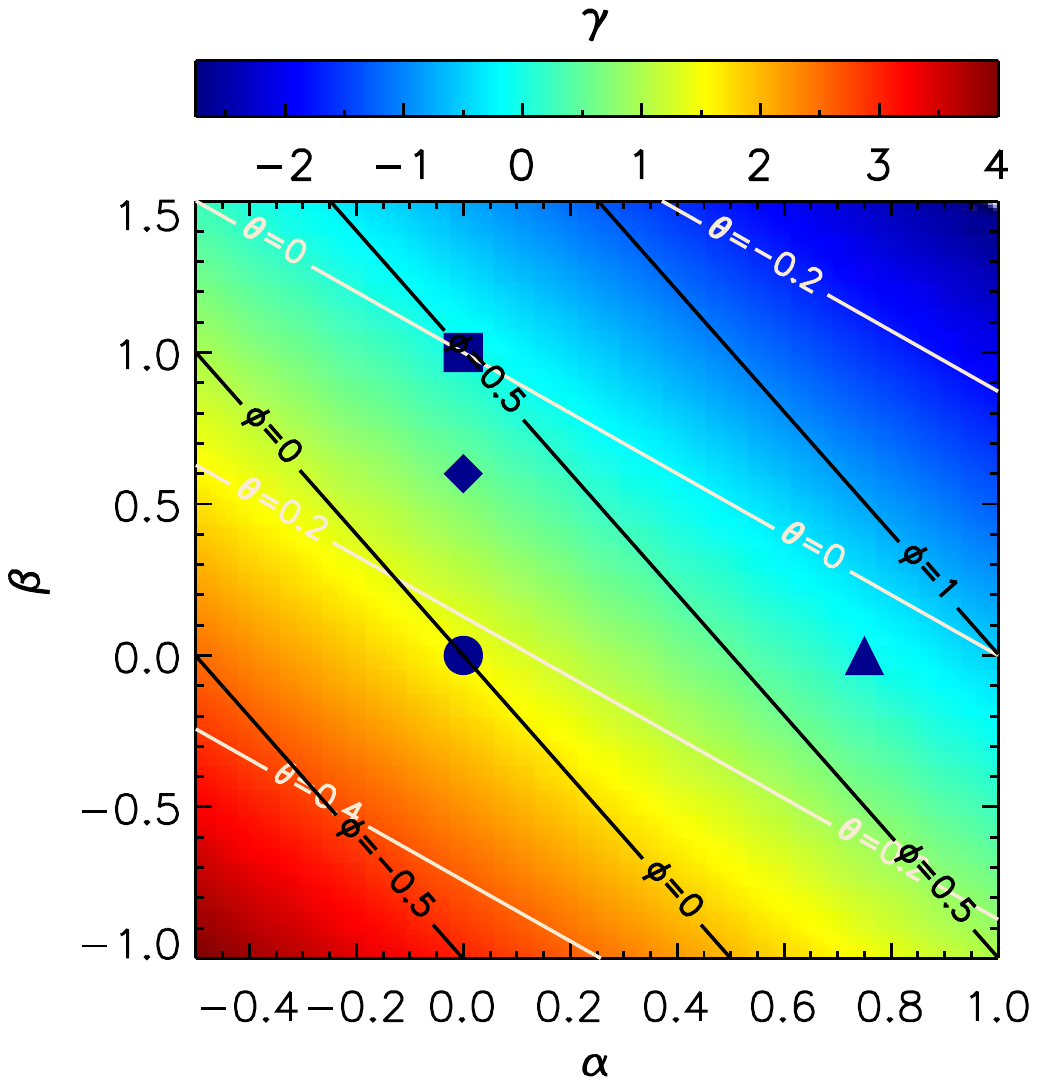}
   \caption{Graphical representation of the Eqs.~\ref{eq:gsl} and \ref{eq:gsl_rel} for the interesting ranges of the values of the exponents $\alpha$ and $\beta$, and the ideal bolometric case ($a_L=1/2$). The symbols correspond to some classical X-ray scaling laws: $M-T$ (dot); $M-M_g$ (square); $M-Y_X$ (diamond); and $M-L$ (triangle).}
\label{fig:rel}
\end{figure}

\begin{figure}[th]
   \centering
   \includegraphics[width=0.5\textwidth,trim=25mm 15mm 25mm 80mm, clip]{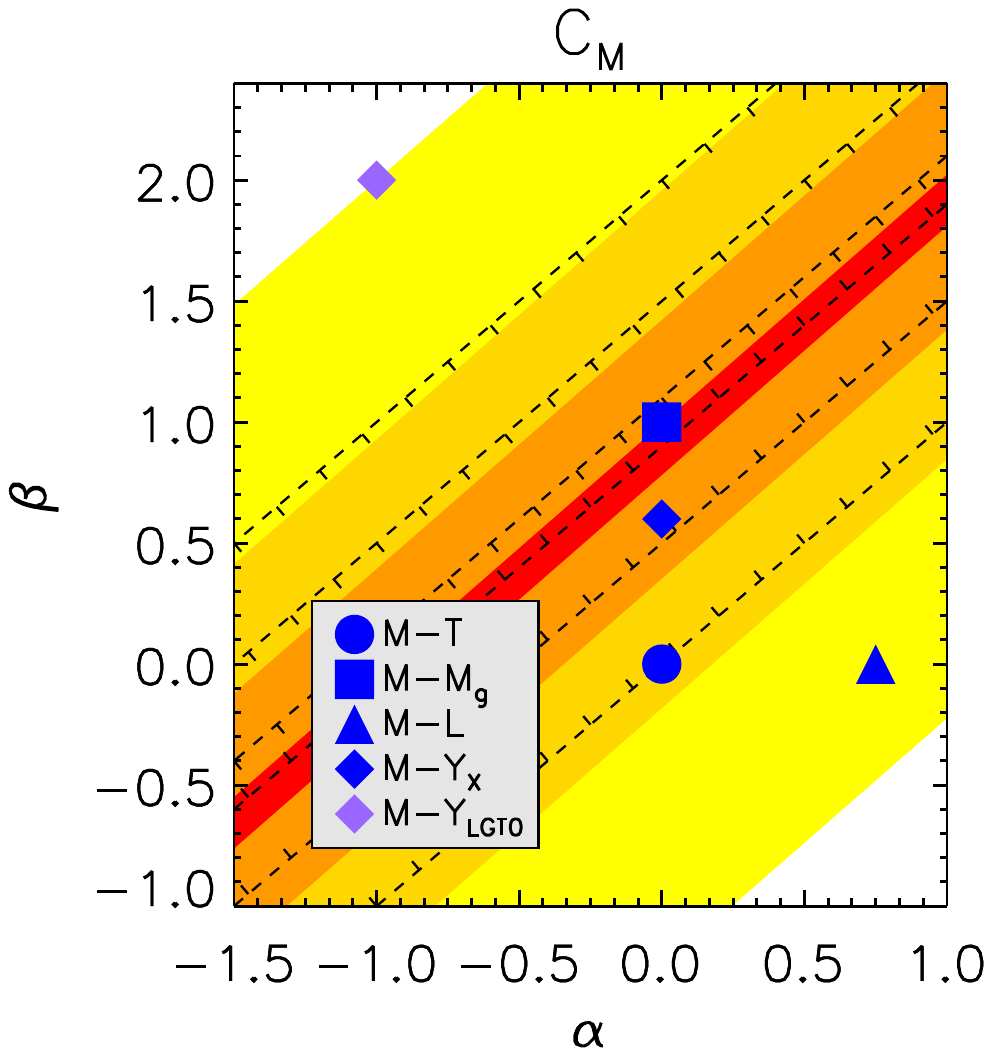}
   \caption{Expected values of $C_M$ (see Eq.~\ref{eq:cm}) for the best-fit values of $\{f_g, f_T\}$ presented in this work (see Eq.~\ref{eq:res}), and assuming $a_L=1/2$. The colored regions indicate the space of the $\alpha-\beta$ plane where $|C_M|<$ 0.1 (red), 0.5 (orange), 1 (dark yellow) and 2 (yellow). The locus of the null evolution is given by the relation $\beta \approx 0.9 +1.0 \alpha$. Dashed contours represent the expectations for a self-similar scenario.}
\label{fig:evol}
\end{figure}

I write this GF here, including the meta-parameters adopted in the present work, $\{f_g, f_T\}$:
\begin{equation}
E_z M  \sim f_T^{\theta} f_g^{-\phi} \; (E_z^{-1} L)^{\alpha} (E_z M_g)^{\beta} T^{\gamma} 
\label{eq:gsl}
,\end{equation}
where the relations
\begin{align}
(3+ & 2 a_L) \, \alpha \, +3 \beta \, +2 \gamma \, =3 \nonumber \\
\theta & = a_L \, \alpha +\gamma \nonumber \\
\phi & = 2 \alpha + \beta
\label{eq:gsl_rel}
\end{align}
among the exponents hold, and $a_L$ indicates the power index of the temperature dependence of the cooling function in different energy bands and yields, for example, $-0.13, -0.12, 0.45$, and $1/2$ in the bands [0.1-2.4] keV, [0.5-2] keV, and [0.01-100] keV, also known as the pseudo-bolometric one, and the ideal bolometric one, respectively.
A graphical representation of these relations is shown in Fig.~\ref{fig:rel}.

These relations are recovered from the scaling between the observables of interest and the mass: $M \sim f_g^{-1} M_g$; $M \sim R \, T_r \sim M^{1/3} T_r \sim T_r^{3/2} \sim (f_T \, T)^{3/2}$, where $T_r$ is the value of reference for a given halo, whereas $T$ is its spectroscopic, observed measurement (see details in Sect.~\ref{sect:ft}); $L \sim n_g^2 \Lambda R^3 \sim M_g^2 T^{a_L} R^{-3} \sim f_g^2 M T^{a_L} \sim f_g^2 f_T^{3/2} T^{3/2+a_L} \sim f_g^2 f_T^{-a_L} M^{1+2/3 a_L}$, $\Lambda$ being the cooling function proportional to $T^{a_L}$. Then, by combining all of them once rearranged as a function of the total mass, $M$, I obtain Eqs.~\ref{eq:gsl} and \ref{eq:gsl_rel}.

For the case $a_L = 1/2$, the relation in Eq.~\ref{eq:gsl_rel} reads $4 \alpha +3 \beta +2 \gamma=3$ and describes a flat plane that intersects the three coordinate axes at specific points, each representing a scaling law with respect to the halo mass through Eq.~\ref{eq:gsl}. By equating different relations, one can then reconstruct any relation between observables. For example, the $M-T$ relation is recovered by imposing the absence of any dependence on $M_g$ and $L$, i.e., $\alpha=\beta=0$; then, $\gamma=3/2$, $\theta=\gamma=3/2$ and $\phi=0$. The $L-T$ relation, instead, is obtained by reconstructing first the $L-M$ relation ($\beta=\gamma=0$; thus, $\alpha = 3/4, \theta = 3/8, \phi = 3/2$) and then by eliminating the mass-dependence in the $M-T$ and $M-L$ relations.

Other relations can be obtained by combining different observables (such as $Y_X = M_{\rm gas} \times T$ -- see \citealt{kravtsov06} -- that I indicate as $Y_{GT}$ being the combination of the gas mass, $G$, and temperature, $T$, for the sake of clarity in the following discussion). The self-similar expectations are $B=\beta=\gamma=3/5$ and $C=-2/5$, with departures propagated through $\theta=\gamma=3/5$ and $\phi=\beta=3/5$. 

Similarly, other relations where two (or more) physical quantities are combined with the same exponent can easily be obtained; for example, (i) $L$ and $T$ ($Y_{LT} = L \times T$) through the conditions $\beta=0$ \, and \, $\alpha=\gamma$, which imply $\alpha=\gamma=1/2$ (from Eq.~\ref{eq:gsl_rel}), self-similar slope $B=1/2$, and evolution $C=-3/2$, and departures $\theta=3/4$ and $\phi=1$, 
(ii) $L$ and $M_g$ ($Y_{LG}$; $\gamma=0$ \, \& \, $\alpha=\beta=3/7$; $B=3/7$, $C=-1$, $\theta=3/14$, $\phi=9/7$), and 
(iii) $L$, $M_g$ and $T$ ($Y_{LGT}$; $\alpha=\beta=\gamma = 1/3$; $B=1/3$, $C=-1$, $\theta=1/2$, $\phi=1$).
It is worth noticing that the two cases $\alpha=0$ and $\beta=0$ identify the loci that minimize the scatter in the mass reconstruction, as discussed in the context of cosmological hydrodynamical simulations in \cite{ettori+12}.

With this model in hand, where the entire departure from self-similarity is ascribed to the meta-parameters $\{f_g, f_T\}$ and a self-consistent redshift evolution can be inferred, 
one can study which scaling relations are less (or more) dependent on the meta-parameters and on $E_z$.
For instance, the dependence on $f_g$ and $f_T$ is canceled by requiring, in  Eq.~\ref{eq:gsl_rel}, that $2\alpha+\beta=0$ (implying $\gamma = 3/2 +3/2\alpha -a_L \alpha$, as for the $M-T$ relation for $\alpha=0$) and $a_L \alpha +\gamma=0$ (i.e. $\alpha+\beta=1$, as for the $M-M_g$ relation with $\alpha=0$), respectively.

As for the redshift evolution, by combining Eqs.~\ref{eq:gsl} and \ref{eq:gsl_rel} with Eqs.~\ref{eq:fg} and \ref{eq:ft}, one can write $M  \sim \, E_z^{C_M}$ with
\begin{equation} 
C_M \, = \, \alpha(a_L \, t_z -2 f_z -1) +\beta (1-f_z) +\gamma \, t_z -1,
\label{eq:cm}
\end{equation}
that reduces to $C_M = -\alpha+\beta-1$ in a pure self-similar scenario.
On the other hand, assuming the best-fit values of the meta-parameters $\{f_g, f_T\}$, I obtain the distribution of $C_M$ values shown in Fig.~\ref{fig:evol}, where a null redshift evolution locus is also identified.
The null-evolution strip does not pass through any of the classical vertices, confirming that building a cosmologically clean mass proxy from a single X-ray observable requires a specific nontrivial combination of $L$, $T$, and $M_g$.
Among the classical scaling relations, $M-M_g$ ($M-L$) is the one with the smallest (largest) expected $ z$ evolution.

\begin{figure}[th]
   \centering
   \includegraphics[width=0.48\textwidth,trim=18mm 20mm 25mm 70mm, clip]{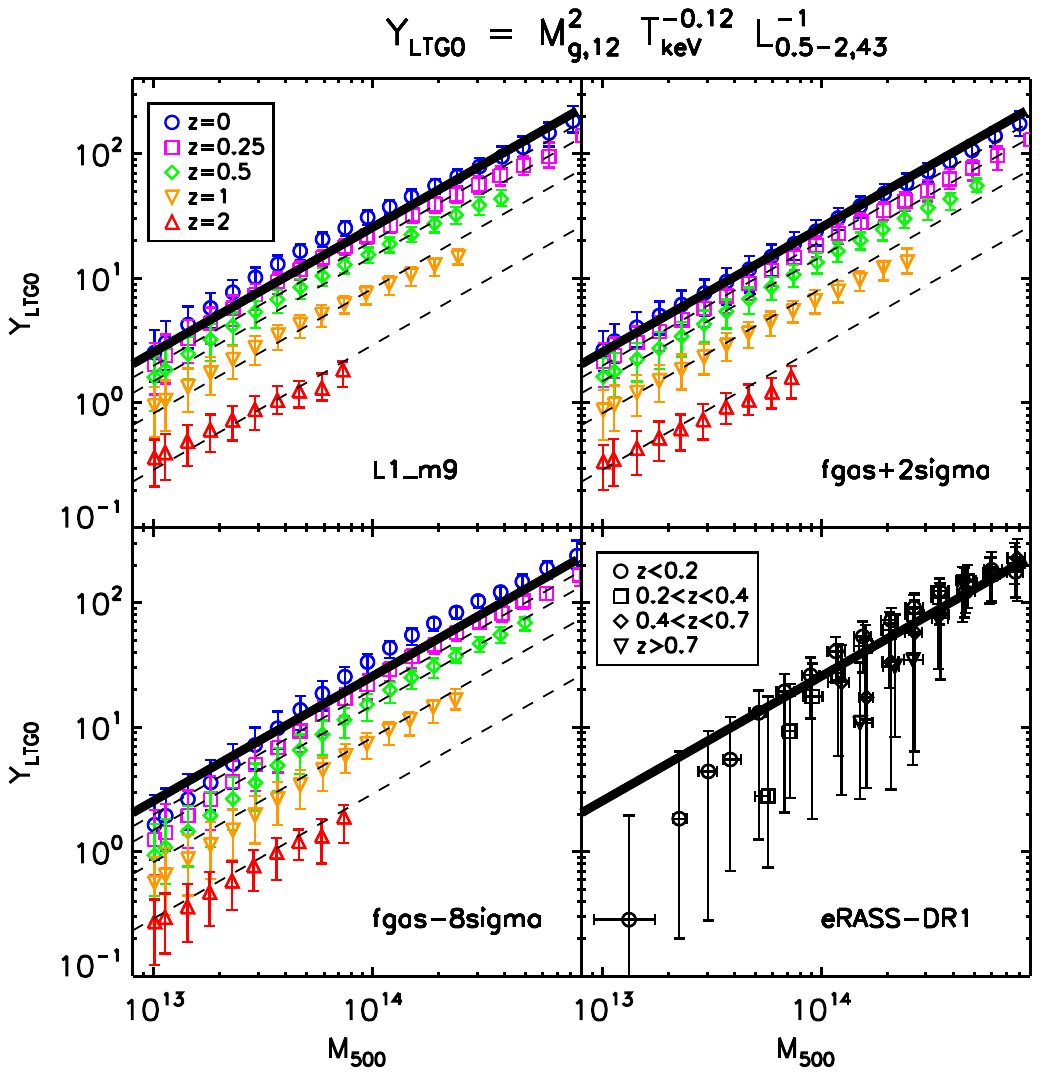}
   \caption{Behavior of the quantity $Y_{LGT0}$ (see Sect.~\ref{sect:proxy}) versus $M_{500}$ and redshift. The exponent $a_L$ for the temperature is set to $-0.12$, appropriate for the analysis in the band $0.5-2$ keV. Sets of FLAMINGO hydrodynamical simulations are used to reconstruct the quantity with different feedback models and at different redshifts; the solid line indicates the local one-to-one correspondence between $Y_{LGT0}$ and mass and represents the relation $Y_{LGT0} = 25.7 \, M_{500, 14}$; dashed lines show the expected $E_z^2$ evolution. In the panel at the bottom right, data extracted from the {\it eROSITA}-DR1 public release are used and corrected by the term $E_z^2$ for comparison with the same local value represented by the solid line. All the data (between $\sim$10k and 99k halos in numerical simulations; 3,388 out of 12,247 in {\it eRASS-DR1} after exclusions of objects with zero values and outside the range in mass and redshift considered here) were binned, and the median and the 16th-84th percentiles were plotted in each bin.
   }
\label{fig:ym}
\end{figure}

\section{The volume proxy $Y_{LGT0}$ and its mass scaling}
\label{sect:proxy}

Imposing no dependence on both $f_g$ and $f_T$ requires $\{\alpha=-1; \beta=2; \gamma=a_L\}$ from Eq.~\ref{eq:gsl_rel}.
This new quantity, $Y_{LGT0} = L^{-1} M_g^2 T^{a_L}$, is thus expected to be independent of the gas mass fraction and gas temperature variation associated with the halo. 
It is worth noticing that it is a direct proxy of the object's volume, as can be easily shown by rearranging the definition of the X-ray bolometric luminosity $L \sim n_g^2 T^{1/2} V \sim M_g^2 T^{1/2} V^{-1}$, implying $Y_{LGT0} \sim V \sim M\,E_z^{-2}$ because the volume scales as $R^3 \sim M/\rho_{\rm c} \propto M E_z^{-2}$.
Therefore, a well-defined redshift evolution, $M \sim Y_{LGT0} E_z^2$ is predicted
(see also Eq.~\ref{eq:cm} and Fig.~\ref{fig:evol}).

I have validated these predictions against measurements estimated both in the FLAMINGO suite of hydrodynamical cosmological simulations \citep{schaye23,helly26}\footnote{for each set of simulations, the halo properties labeled as {\it TotalMass, HotGasMass, SpectroscopicLikeTemperatureWithoutRecentAGNHeating, XRayLuminosityWithoutRecentAGNHeating} were saved and downloaded from \url{https://flamingo.strw.leidenuniv.nl/} for the analysis}
and in the analysis of the {\it eROSITA} X-ray selected galaxy groups and clusters \citep{bulbul24}\footnote{using the values tabulated in the columns {\it M500, Mgas500, L500\_0520, KT} of the file {\tt erass1cl\_primary\_v3.2.fits} available at \url{https://erosita.mpe.mpg.de/dr1/AllSkySurveyData_dr1/Catalogues_dr1/}}.
The total masses drawn from the {\it eROSITA} catalog, where they are calibrated versus weak-lensing measurements, and from the FLAMINGO suite of cosmological simulations are taken as true masses by construction.

I show the reconstructed quantity $Y_{LGT0}$ as a function of mass and redshift in Fig.~\ref{fig:ym}. When compared with the values expected under different feedback schemes (the reference one ``L1\_m9'' and some extreme modifications of it, ``fgas+2sigma'' and ``fgas-8sigma''), no significant variations are noted, even in the redshift evolution.
I reach a similar conclusion when I reconstruct the observed quantity $Y_{LGT0}$ from the products of the {\it eROSITA} analysis, after rescaling by the expected evolution, $E_z^2$. A possible mismatch at masses lower than $10^{14} M_{\odot}$ is present, even though this mass range is potentially more affected by observational systematic uncertainties; nonetheless, there is still marginal consistency between the observed distribution and the predicted behavior.

Practically, to measure $Y_{LGT0}$, every quantity — $L$, $T$, and $M_g$ — must be integrated or extracted within $R_{500}$, but $R_{500}$ itself is defined implicitly through the mass, $M_{500}$, which is what we are trying to determine. 
So, the standard iterative scheme also used to evaluate, for example, $Y_X = M_g T$, is recommended. 
We started with an initial guess, $R_{500}^{(0)}$ (e.g., from a rough $L$–$M$ or $T$–$M$ prior); then, from the gas density and luminosity profiles, we measured $L$ and $M_g$, and by fitting a spectrum integrated over the same region, $T$, allowing us to compute $Y_{LGT0}^{(n)}$. We update the mass, $M_{500}^{(n+1)} = A_Y \, Y_{LGT0}^{(n)} \, E_z^2$ (see Fig.~\ref{fig:ym}), and the corresponding radius, $R_{500}^{(n+1)} = \left(3M_{500}^{(n+1)}/4\pi\,500\,\rho_c\right)^{1/3}$, and repeated until $|R_{500}^{(n+1)} - R_{500}^{(n)}|/R_{500}^{(n)} < \epsilon$ (typically 1–2\%).
This procedure is expected to converge robustly and rapidly because of the linearity of $M \propto Y_{LGT0}$ and despite the possible presence of small deviations from the pure power law due to, for example, profile shapes, cool cores, and a high sensitivity (higher than for $Y_X$ by roughly a factor of 2 for the different dependence on $M_g$, for instance) to the aperture error. In particular, to avoid any cool-core bias, it is standard practice to measure the temperature, $T$, in an annulus that excludes the core (typically $[0.15$–$1]\,R_{500}$), so the spectral extraction region itself changes between iterations.

\section{A meta-analysis}
\label{sect:ana}

For each of the scaling laws under consideration, I constructed the expected slope, $B$, and the redshift evolution, $C$, starting from a self-similar value and adding the contributions from the four parameters that are the only free parameters in the analysis (see App.~\ref{app:mod}). The advantages of this meta-analysis are manifold: first, a dimensionality reduction of the problem from thousands of data points into a few well-defined sets of a few (meta-)parameters; then, the adoption of physically motivated (meta-)parameters that enrich and extend the self-similar model; furthermore, this meta-analysis provides a framework in which a consistency check can be applied in such a way that outliers in the published best-fit parameters can be recognized and investigated.
On the other hand, critical aspects of the statistical framework are also evident, such as the neglect of covariance propagation among the quantities considered and of relative errors.

 \begin{figure}[htb]
   \centering
   \includegraphics[width=0.5\textwidth,trim=22mm 130mm 10mm 25mm, clip]{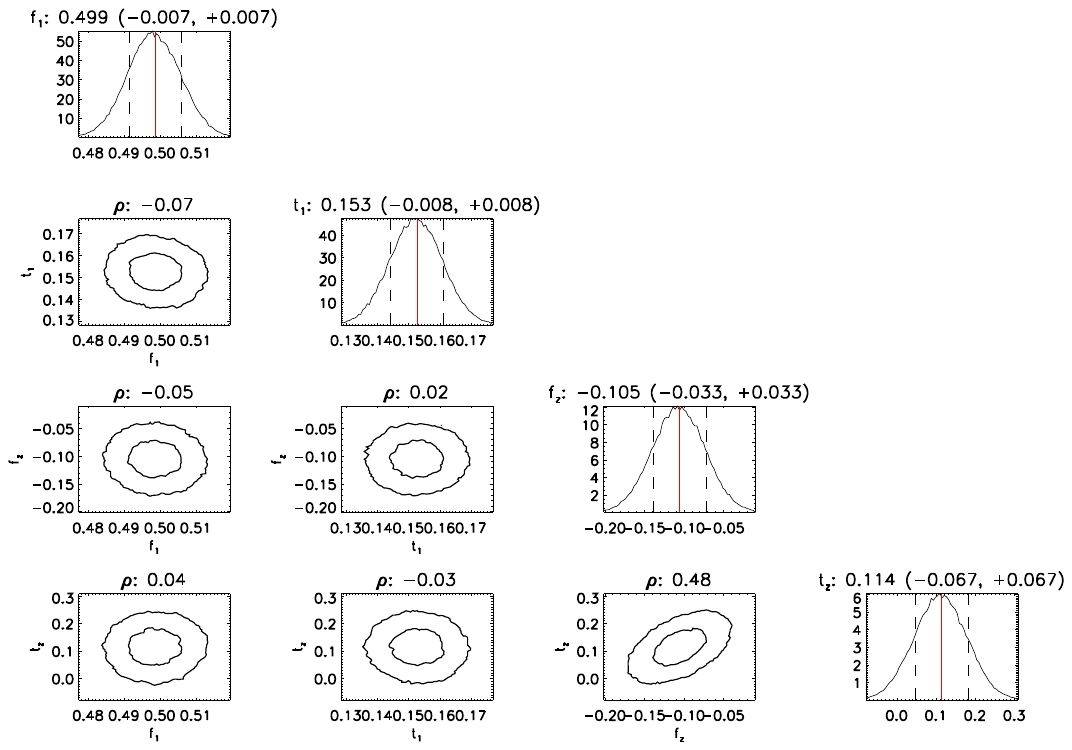}
      \caption{Corner plot of the likelihood distribution of the four parameters, describing the departure from self-similarity, after an MCMC analysis with 64 walkers, 20,000 iterations, and 5,000 burn-in steps. Uniform priors (in the range $[0, 1]$ for $f_1$ and $t_1$; $[-1, 1]$ for $f_z$ and $t_z$) are assumed. The contours refer to the 1 and 2 $\sigma$ confidence levels for two parameters; the dashed lines indicate 1 $\sigma$ errors for a single parameter.}
\label{fig:corner}
\end{figure}

\begin{figure}[ht]
   \centering
   \includegraphics[width=0.5\textwidth,trim=25mm 15mm 20mm 80mm, clip]{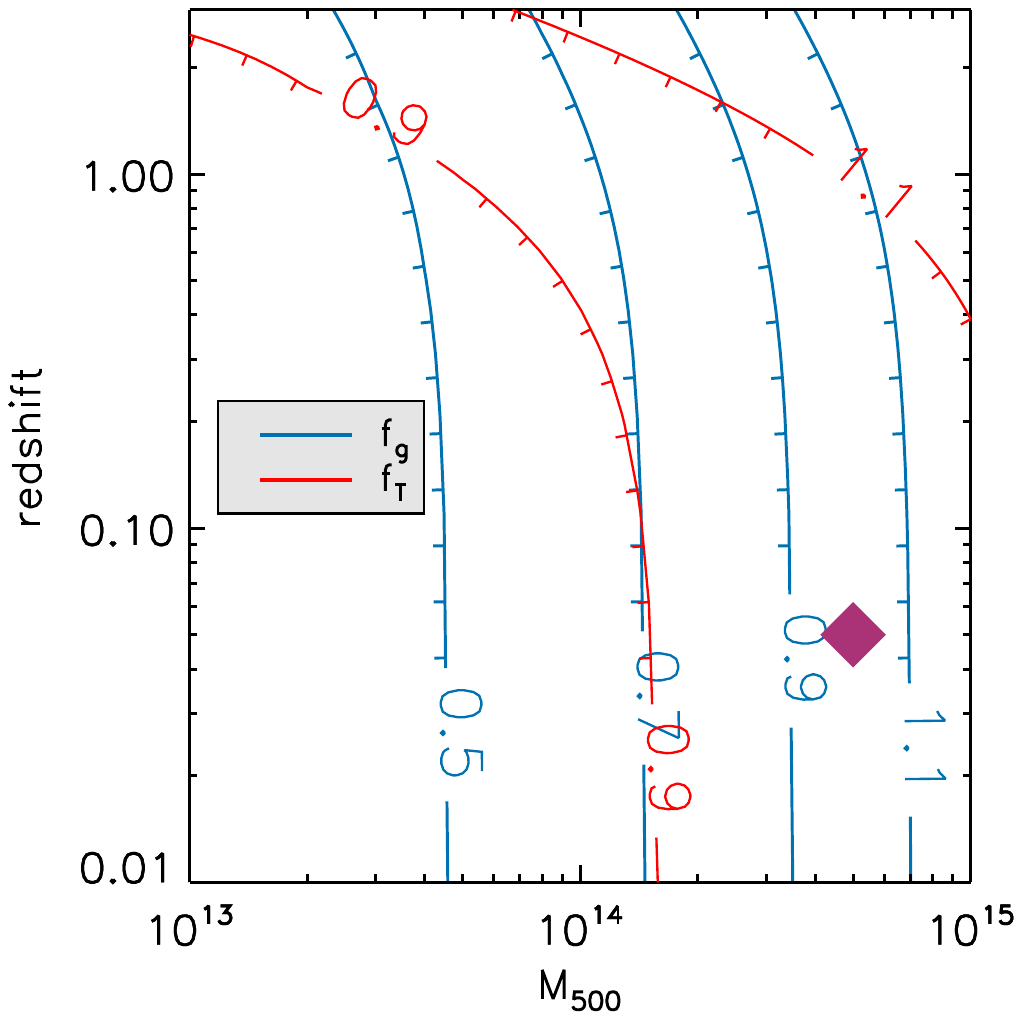}
      \caption{Variations in $\{f_g, f_T\}$ as a function of halo mass and redshift. The changes refer to a value of 1 assumed at $\{M, z\} = \{5 \times 10^{14} M_{\odot}, 0.05\}$.}
\label{fig:ftfg}
\end{figure}

Using an MCMC analysis (via {\tt emcee} in IDL), I constrained the four (meta-)parameters (and their correlations; see the corner plot in Fig.~\ref{fig:corner}) by reproducing the best-fit values reported in the recent literature and spanning a wide range in mass and redshift (see App.~\ref{app:pub}).
I obtained the following results:
\begin{align}
f_1 = & 0.499 \ (\pm 0.007); \; f_z = -0.105 \ (\pm 0.033) \nonumber \\
t_1 = & 0.153 \ (\pm 0.008); \; t_z = 0.114 \ (\pm 0.067).
\label{eq:res}
\end{align}
Some covariance is present between the parameters describing the $z$ evolution, $f_z$ and $t_z$, as a consequence of the limited number of datasets covering the regime at high redshift.
How $f_g$ and $f_T$ change as a function of mass and redshift is shown in Fig.~\ref{fig:ftfg}, with a larger relative impact due to the halo mass than the redshift.

\begin{figure}[ht]
\centering \includegraphics[width=0.5\textwidth,trim=25mm 15mm 20mm 80mm, clip]{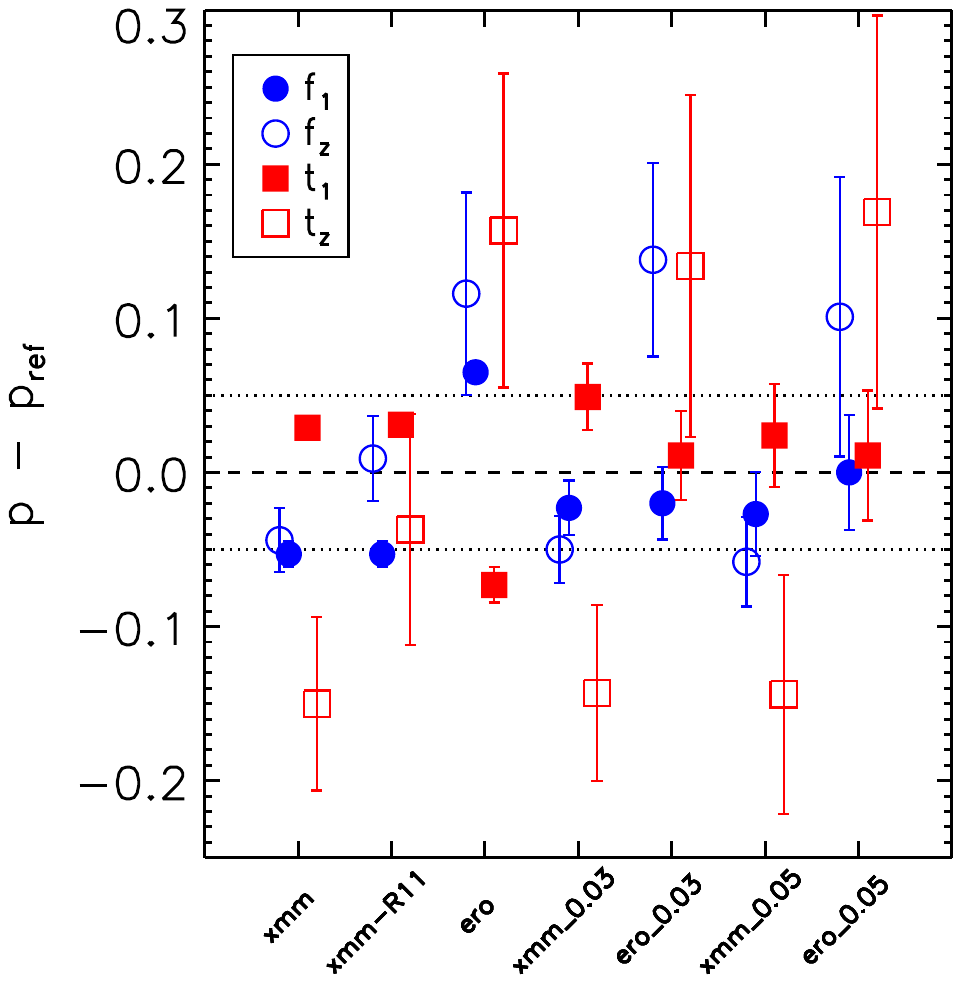}
\caption{Comparison between the reference values in Eq.~\ref{eq:res} (“$p_{\rm ref}$”) and the best-fit results obtained for different subsamples (from left to right): scaling relations obtained by XMM measurements only (“xmm”); the same excluding the relations in \cite{reichert11} (“xmm-R11”) showing that they matter only for the evolution terms; using relations from eROSITA analysis only (“ero”); then, the previous ones fixing the relative uncertainties to be larger than 3\% (“\_0.03”) and 5\% (“\_0.05”). The error bars indicate the difference in quadrature between the errors on $p$ and $p_{\rm ref}$, because the subsample is part of the entire sample analyzed. Dotted lines show differences of $\pm 0.05$. }
\label{fig:compa}
\end{figure}

To assess robustness against inter-telescope cross-calibration systematics, I separated the 39 constraints into {\it XMM-Newton}-based (23 relations, plus four more obtained from a mixture of data but dominated by {\it XMM-Newton} ones) and {\it eROSITA}-based (12 relations) subsamples, and repeated the MCMC analysis independently. 
I plot the results for the four meta-parameters against the best-fit values in Eq.~\ref{eq:res} in Fig.~\ref{fig:compa}. 
Whereas the evolution terms are mutually consistent within 1-2 $\sigma$ and agree with the full-sample values in Eq.~\ref{eq:res}, the slopes differ by roughly $\pm0.05$, assessing what the current residual calibration offsets are \citep[see e.g.][]{turner22,migkas24}; a homogeneously calibrated sample will sharpen this test.

Overall, the best-fit results are quite consistent with the ones discussed in E23,
with an increase of 10--25\% on the slopes, and slightly more evident negative evolution in $f_g$ and less clear positive evolution in $f_T$.

\begin{figure*}[ht]
   \centering
   \hbox{ 
\includegraphics[page=1,width=0.33\textwidth,trim=30mm 20mm 10mm 70mm, clip]{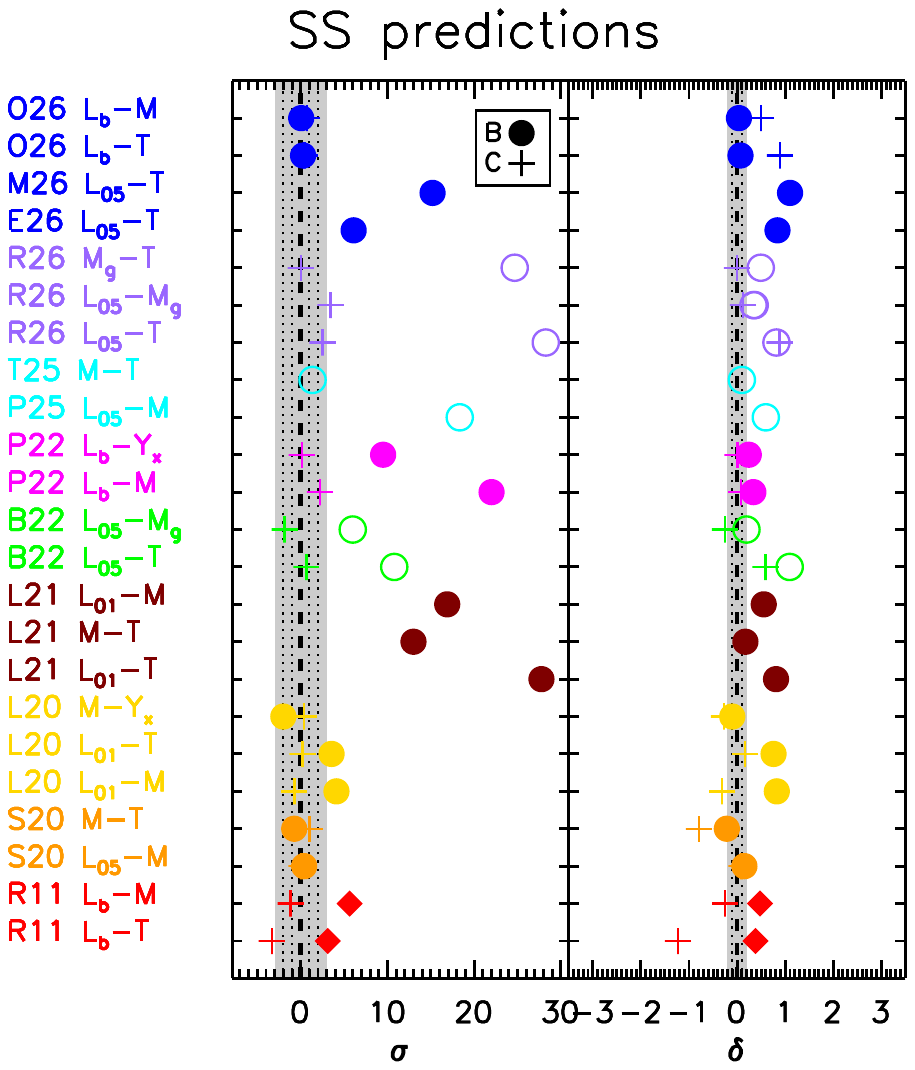}   \includegraphics[page=1,width=0.33\textwidth,trim=30mm 20mm 10mm 70mm, clip]{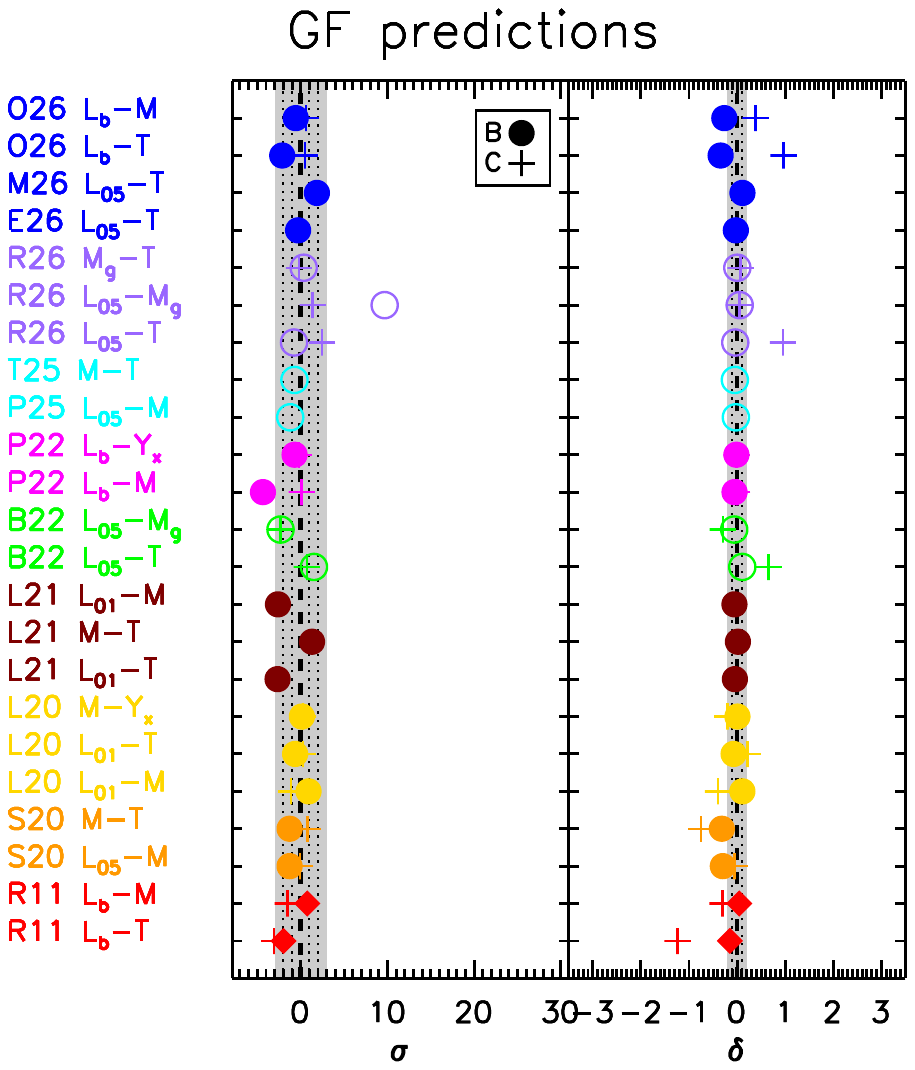}
\includegraphics[page=2,width=0.33\textwidth,trim=30mm 20mm 10mm 70mm, clip]{figures/pred_e26.pdf} 
    } \caption{Deviations in terms of $\sigma = (O-M)/(\epsilon_O^2 +\epsilon_M^2)^{1/2}$ and $\delta = (O-M)/M$, where $O$ and $M$ represent the values from observation and model, respectively, for the slope $B$ (circles; filled: {\it XMM-Newton} data; open: {\it eROSITA} data; filled diamond: mostly {\it XMM-Newton} and {\it Chandra} data, with prevalence of the former) and $E_z$-evolution $C$ (crosses; see Eq.~\ref{eq:scalaw}). The shaded regions identify intervals between -3 and 3 $\sigma$ and between -0.2 and 0.2 $\delta$, respectively. The panels on the left and at the center, referring to the self-similar (SS) scenario and the current analysis, respectively, use the same X axis intervals to emphasize the range of differences implied by SS expectations that the model described here can accommodate. 
    The panel on the right is a zoomed-in version of the central panel. }
\label{fig:SS_rev}
\end{figure*}

In Fig.~\ref{fig:SS_rev}, I show the improvement (in terms of both the standard deviation, $\sigma$, with respect to the sum in quadrature of the measured error and the uncertainties quoted in Eq.~\ref{eq:res}, and bias, $\delta$, in the central value) from a scenario in which no departures are considered to one in which they are modeled with $\{f_g, f_T\}$.
Among 39 relations considered in this meta-analysis (23 on the value of $B$, and 16 on the value of $C$), 19 (14) have a tension $>3 \sigma$ ($>5 \sigma$) when self-similarity is adopted. 
These numbers decrease to 3 (1) when $f_g - f_T$ are left free to vary.
These three relations are, in order of significance in tension: the $L_{05} - M_{\rm g}$ relation in \cite{ramos-ceja26} ($+9.7 \sigma$); $L_b-M$ relation in \cite{pratt22} ($-4.4 \sigma$); and the redshift evolution in the $L_b-T$ relation in \cite{reichert11} ($-3.1 \sigma$).
It is worth noting that the most significant outlier in the GF analysis -- the slope of $L_{05} - M_{\rm g}$ in \cite{ramos-ceja26} -- also carries the smallest relative statistical error among all constraints considered here (0.5\% at $1 \sigma$), far below any plausible systematic floor. As a sensitivity test, I imposed a minimum relative statistical error of 3\% on all constraints and repeated the MCMC. The choice of 3\% reflects an assumed approximation of the level of known systematic floors (e.g., due to the absolute calibration of the observed quantities, selection effects), and is not a derived quantity. 
The best-fit parameters remain highly consistent with Eq.~\ref{eq:res} ($f_1 = 0.48 \pm 0.02$, $f_z = -0.10 \pm 0.03$, $t_1 = 0.19 \pm 0.02$, $t_z = 0.10 \pm 0.07$), and the number of relations in tension at $> 3\sigma$ falls to two (and none above $5\sigma$). Furthermore, when the minimum relative statistical error is raised to 5\%, all tensions between the published and best-fit values fall below $3\sigma$.
I stress that this is a diagnostic of error underestimation in the literature, not evidence of improved physical modeling.

An interesting implication of this study is the modeling of the gas mass fraction.
From the adopted scaling and the best-fit results in Eq.~\ref{eq:res}, I obtain $f_g \propto M^{2 f_1/(3+3t_1)} \, E_z^{f_z +(2-3t_z)/(3+3t_1)} \sim M^{0.29 (\pm 0.01)} \, E_z^{0.13 (\pm 0.04)}$. 
These constraints can be directly compared with results published in recent literature on the observed slope of the $f_g-M_{500}$ relation, showing: consistency with the estimates of $0.26 \pm 0.03$ \citep{gonzalez+13}, $0.22 \pm 0.05$ \citep{eckert21}, and $0.21 \pm 0.13$ \citep{akino22} in (mostly) X-ray selected nearby objects; still a good agreement with the distribution of the 91 objects selected from the SPT survey and followed-up with the X-ray telescope {\it Chandra} in the redshift range 0.2--1.25, $f_g \sim M_{500}^{0.32 \pm 0.07} Ez^{-0.17 \pm 0.15}$ \citep{chiu18}; and probably a slight tension with the range of values between $0.38$ and $0.5$ in different redshift bins and up to $z\sim 1.1$  as measured in 10,440 objects selected in the {\it eROSITA} eRASS-1 survey \citep{bulbul24}, and with the value of $0.39 \pm 0.02$ obtained for optically (GAMA) selected local objects followed-up with {\it eROSITA} signal in \cite{popesso26}.

\section{Summary and conclusions}

This work describes how the ``generalized form'' GF contracts all X-ray cluster scaling laws into one equation,
$E_z M ~ f_T^\theta f_g^{-\phi} (E_z^{-1} L)^\alpha (E_z M_g)^\beta T^\gamma$,
with the constraint $(3+2a_L) \alpha + 3 \beta + 2 \gamma = 3$, and $\theta = a_L \alpha + \gamma$, $\phi = 2 \alpha + \beta$. The entire departure from self-similarity is encoded in two phenomenological power laws: $f_g = f_0 T^{f_1} E_z^{f_z}$, the gas mass fraction varying with mass and redshift; and $f_T = t_0 T^{t_1} E_z^{t_z}$, the temperature offset between the reference and spectroscopic values.
The physical meaning of $f_T$ (described in detail in  Sect.~\ref{sect:ft}) encompasses at least three contributions: the radial temperature gradient of the ICM and the projection effects (the primary motivations), and the thermalization state of the gas, which depends on the system's dynamical age. The calibrated parameters $t_1$ and $t_z$ therefore reflect a combination of structural, dynamical, and observational effects.

The GF framework reduces the degrees of freedom from $O(N_{\rm relations} \times 2)$ slope+evolution pairs to just the four meta-parameters $\{f_1, f_z, t_1, t_z\}$; it provides a self-consistent framework for predicting any scaling relation once those four numbers are calibrated; and it cleanly identifies combinations (such as $Y_{LGT0}$) that are immune to both meta-parameters.
In particular, I present an extension and an update of the analysis discussed in E23, by investigating, with a new and large set of published scaling relations among integrated quantities describing the gas in halos of mass $M_{500}> 10^{13} M_{\odot}$ and up to redshift 1.5, how the departures from the self-similar scenario can be interpreted in this GF framework.

Using 39 published constraints (23 on the value of the slope $B$, and 16 on the exponent $C$ of the redshift evolution traced with $E_z$; 
23 relations based on {\it XMM-Newton} observations, 12 on {\it eROSITA-DR1} data, and 4 from a sample observed from different X-ray telescopes but with a prevalence of data collected with {\it XMM};
see Table~\ref{tab:pub}), I calibrate the (meta-)parameters $\{f_1, f_z; t_1, t_z\}$ (see description of the procedure implemented and results obtained in Sect.~\ref{sect:ana}), and obtain a stronger $T$ dependence in $f_g$ (that translates into a mass dependence well consistent with some recent studies) and a mild (different from zero at $\sim$ 2-3 $\sigma$ level of confidence) redshift evolution for both.
As I discuss in Sect.~\ref{sect:ana}, the best-fit results still depend on possible residual systematic uncertainties related to, for example, the cross-calibration of different instruments, and to some underestimates of the statistical error in published summary statistics. I quantify these light dependencies at about $\pm 0.05$ on the central value of $f_1$ and $t_1$. The large statistical error associated with the terms describing the redshift evolution is expected to still dominate their error budget.

Of these 39 published constraints, 19 (14) exhibit tensions $>3 \sigma$ ($>5 \sigma$) when self-similarity is assumed. 
These numbers decrease to 3 (1) when $\{f_g, f_T\}$ are left free to vary, which likely indicates anomalies in the published values. For instance, by imposing a relative statistical error not lower than 3 (5) percent on the constraints extracted from the literature, the number of values in tension at $>5\sigma$ drops to zero, and those still at $>3\sigma$ are reduced to two (none), with negligible variations in the best-fit estimates of $\{f_1, f_z; t_1, t_z\}$.

By propagating these best-fit values through the GF of the X-ray scaling relations originally presented in \cite{ettori15}, I make solid predictions about both the expected slope and redshift evolution of several X-ray scaling laws (see Sect.~\ref{app:mod}). 
The GF framework permits the study of interesting properties of these relations, such as: (i) the expected self-similar behavior and its modified version through the meta-parameters $\{f_g, f_T\}$, of the relations that combine two (or more) physical quantities (such as $Y_X$); (ii) the locus where the redshift evolution with respect to the mass is minimized ($\beta \approx 0.9+1.0 \alpha$); 
(iii) the construction of a new quantity $Y_{LGT0} = L^{-1} M_g^2 T^{a_L}$, 
which is a proxy for the cluster's volume, does not depend on $f_g$ and $f_T$ by construction and is predicted to relate directly to the mass via a well-defined redshift evolution: $M \sim Y_{LGT0} f_g^0 f_T^0 E_z^2$ (see Sect.~\ref{sect:proxy}).
Furthermore, this calibration allows us to properly rescale the thermodynamic quantities when constructing the ``universal'' profiles (see, e.g., \citealt{arnaud10}; \citealt{ettori20}; E23), accounting for their dependence on halo mass (or temperature) and redshift (see Table~\ref{tab:norm}).

I note that the GF framework can be expanded to include, for example, 
(i) other multiwavelength observables, such as the SZ Compton-y parameter and weak-lensing mass \citep[see][]{ettori15};
(ii) a hydrostatic mass bias dependent on temperature and redshift \citep[see][for a treatment of it as a constant]{ettori23}; 
(iii) a second-order behavior relaxing the assumption of constant meta-parameters, for example in the form $f_1(T) = f_{1,0} + f_{1,1} \log T; f_z(T) = f_{z,0} + f_{z,1} \log T$; 
(iv) the scatter around the predicted GF mean relations, with also a dependence on $f_g$ and $f_T$ through their halo-to-halo variance encoding physical information about assembly history.
Regarding the last item, I note that the expected intrinsic scatter associated with $Y_{LGT0}$ is nonzero despite its zero leading-order dependence on $f_g$ and $f_T$; rather, it reflects contributions from halo-to-halo variations in gas structure, luminosity, gas mass, and temperature, with correlations among them impacting the overall variance.
Moreover, a forward-modeled GF allows for the direct use of cluster catalogs — rather than summary statistics, as in the meta-analysis I present here, which treats published slopes as unbiased estimates — by propagating any selection function (such as a flux limit or an SZ signal-to-noise cut) directly into the likelihood.

The new datasets that are becoming available, accurately selected and homogeneously exposed to X-rays, such as CHEX-MATE \citep{chexmate21}\footnote{\protect\url{http://xmm-heritage.oas.inaf.it/}} and X-GAP \citep{eckert24}\footnote{\protect\url{https://www.astro.unige.ch/xgap/front-page}}, will permit detailed studies of both the relations among integrated quantities and the universal behavior of spatially resolved gas thermodynamic properties.
Complementary analyses linking the slopes and evolution of the reconstructed X-ray scaling laws in numerical simulations to the meta-parameters will enable us to assess and constrain the level of feedback acting on the baryon energy and distribution as a function of halo mass and cosmic time.

\begin{acknowledgements}
I thank Y. Emre Bahar, Vittorio Ghirardini, Lorenzo Lovisari, Thomas Reiprich, and Mauro Sereno for comments on the manuscript. 
I am grateful to Joey Braspenning for advice on downloading FLAMINGO data, and to the anonymous referee for their thoughtful and constructive remarks, which strengthened the paper.
I acknowledge the financial contribution from {\it Theory Grant / Bando INAF per la Ricerca Fondamentale 2024} on ``Constraining the non-thermal pressure in galaxy clusters with high-resolution X-ray spectroscopy'' (1.05.24.05.10).
All the files used in this study are available upon reasonable request.
\end{acknowledgements}

\bibliographystyle{aa} 
\bibliography{icmz}

\begin{appendix}
\nolinenumbers
\section{Relations investigated}

I list here both the theoretical predictions and the published values used in the present work for the slopes $B$ and the redshift evolution $C$.

\subsection{Theoretical relations}
\label{app:mod}

The model that describes departures from self-similarity is detailed in Sect.~\ref{sect:gen} (see also Sects.~2.1 and 2.2 in E23).
Here, I describe how the slopes $B$ and redshift-evolution $C$ of the scaling relations studied in this work are modified by the four parameters $\{f_1, f_z; t_1, t_z\}$. The self-similar expectations are obtained by fixing them to $0$.

\begin{description}
    \item[{\boldmath{${M-T}$}} ($\alpha=\beta=0$):] 
    $B = 3/2+3/2 \cdot t_1; \\
    C  = 3/2 \cdot t_z -1$
    \item[{\boldmath{$M_g - T$}} ($\alpha=\beta=0$ \, \& \, $\alpha=\gamma=0$):] $B = 3/2 \cdot (1+t_1)+f_1; \\
    C = f_z +3/2 \cdot t_z-1$
    \item[{\boldmath{$M-Y_X$}} ($\alpha=0$ \, \& \, $\beta=\gamma$):] $B = 3 \cdot (1+t_1)/(5 +3 \cdot t_1 +2 \cdot f_1); \\
    C = -(f_z \cdot (1+t_1)+(1+f_1) \cdot (2 -3 \cdot t_z)) / ( 5 +3 \cdot t_1 +2 \cdot f_1 )$
    \item[{\boldmath{$L-T$}} ($\alpha=\beta=0$ \, \& \, $\beta=\gamma=0$):] $B = 3/2+a_L +3/2 \cdot t_1+2 \cdot f_1; \\
    C = 1+2 \cdot f_z +3/2 \cdot t_z$
    \item[{\boldmath{$L-M$}} ($\beta=\gamma=0$):] $B = (1+2/3 \cdot a_L+t_1+4/3 \cdot f_1)/(1+t_1); \\
    C = 1+2 \cdot f_z+3/2 \cdot t_z+(2-3 \cdot t_z) \cdot (1/2 +a_L/3 +t_1/2 +2/3 \cdot f_1)/(1+t_1) $
    \item[{\boldmath{$L-M_g$}} ($\alpha=\gamma=0$ \, \& \, $\beta=\gamma=0$):] $B = (3+2 \cdot a_L +3 \cdot t_1+4 \cdot f_1)/(3 +3 \cdot t_1+2 \cdot f_1); \\
    C = (3+2 \cdot a_L +3 \cdot t_1+4 \cdot f_1)/(3 +3 \cdot t_1+2 \cdot f_1) \cdot (1 -f_z -3/2 \cdot t_z) +1 +2 \cdot f_z +3/2 \cdot t_z$.
  \end{description}

In Table~\ref{tab:bc}, I quote the expected values of the slope $B$ and redshift evolution $C$ for the most relevant relations.

\begin{table}[hbt]
   \caption{Best-fit values for the quoted relations.}
\begin{center} \begin{tabular}{ccc} \hline
 Rel.  &   $B$  &  $C$  \\  \hline
MT & $1.730 \, (0.012) \, [1.500]$ & $-0.829 \, (0.101) \, [-1.000]$ \\
MgT & $2.229 \, (0.014) \, [1.500]$ & $-0.934 \, (0.106) \, [-1.000]$ \\
MYx & $0.536 \, (0.008) \, [0.600]$ & $-0.366 \, (0.047) \, [-0.400]$ \\
LbT & $3.178 \, (0.018) \, [1.950]$ & $0.961 \, (0.120) \, [1.000]$ \\
L1T & $2.597 \, (0.018) \, [1.370]$ & $0.961 \, (0.120) \, [1.000]$ \\
L5T & $2.608 \, (0.120) \, [1.380]$ & $0.961 \, (0.120) \, [1.000]$ \\
LbM & $1.837 \, (0.010) \, [1.300]$ & $2.484 \, (0.107) \, [2.300]$ \\
L1M & $1.502 \, (0.009) \, [0.913]$ & $2.206 \, (0.083) \, [1.913]$ \\
L5M & $1.508 \, (0.009) \, [0.920]$ & $2.211 \, (0.084) \, [1.920]$ \\
LbMg & $1.426 \, (0.003) \, [1.300]$ & $2.293 \, (0.021) \, [2.300]$ \\
L1Mg & $1.166 \, (0.003) \, [0.913]$ & $2.050 \, (0.032) \, [1.913]$ \\
L5Mg & $1.170 \, (0.003) \, [0.920]$ & $2.054 \, (0.032) \, [1.920]$ \\
\hline \end{tabular} \end{center}
   \label{tab:bc}
   \tablefoot{The corresponding self-similar expectations are indicated in the square bracket. }
\end{table}

\subsection{Published results}
\label{app:pub}

In Table~\ref{tab:pub}, I present the list of the published results used to calibrate the meta-parameters $\{f_g, f_T\}$.

\begin{table}
    \caption{Published relations considered in the present work.}
\setlength\tabcolsep{2.2pt} \begin{center} \begin{tabular}{cccccc} \hline
Ref. & Par. & value & $N_{\rm obj}$ & $\Delta M$ ($10^{14} M_{\odot}$) & $\Delta z$ \\ \hline
R11 & $B_{LbT}$ & $2.70 \pm 0.24$ & 232 & 0.5 - 20.0 & 0.30 - 1.46 \\
R11 & $C_{LbT}$ & $-0.23 \pm 0.37$ &  -  &  -  &  -  \\
R11 & $B_{LbM}$ & $1.92 \pm 0.11$ &  -  &  -  &  -  \\
R11 & $C_{LbM}$ & $1.73 \pm 0.49$ &  -  &  -  &  -  \\
S20 & $B_{L5M}$ & $1.06 \pm 0.35$ & 105 & 0.1 - 10.0 & 0.05 - 1.10 \\
S20 & $C_{L5M}$ & $2.10 \pm 2.08$ &  -  &  -  &  -  \\
S20 & $B_{MT}$ & $1.18 \pm 0.43$ &  -  &  -  &  -  \\
S20 & $C_{MT}$ & $-0.21 \pm 0.78$ &  -  &  -  &  -  \\
L20 & $B_{L1M}$ & $1.67 \pm 0.18$ & 120 & 2.4 - 17.6 & 0.06 - 0.55 \\
L20 & $C_{L1M}$ & $1.33 \pm 0.80$ &  -  &  -  &  -  \\
L20 & $B_{L1T}$ & $2.41 \pm 0.29$ &  -  &  -  &  -  \\
L20 & $C_{L1T}$ & $1.17 \pm 0.82$ &  -  &  -  &  -  \\
L20 & $B_{MYx}$ & $0.54 \pm 0.03$ &  -  &  -  &  -  \\
L20 & $C_{MYx}$ & $-0.29 \pm 0.29$ &  -  &  -  &  -  \\
L21 & $B_{L1T}$ & $2.48 \pm 0.04$ & 40 & 0.1 - 10.0 & 0.01 - 0.80 \\
L21 & $B_{MT}$ & $1.76 \pm 0.02$ &  -  &  -  &  -  \\
L21 & $B_{L1M}$ & $1.42 \pm 0.03$ &  -  &  -  &  -  \\
B22 & $B_{L5T}$ & $2.89 \pm 0.14$ & 265 & 0.1 - 7.8 & 0.02 - 0.94 \\
B22 & $C_{L5T}$ & $1.59 \pm 0.90$ &  -  &  -  &  -  \\
B22 & $B_{L5Mg}$ & $1.10 \pm 0.03$ &  -  &  -  &  -  \\
B22 & $C_{L5Mg}$ & $1.44 \pm 0.26$ &  -  &  -  &  -  \\
P22 & $B_{LbM}$ & $1.74 \pm 0.02$ & 93 & 0.5 - 20.0 & 0.05 - 1.13 \\
P22 & $C_{LbM}$ & $2.50 \pm 0.09$ &  -  &  -  &  -  \\
P22 & $B_{LbYx}$ & $0.97 \pm 0.02$ &  -  &  -  &  -  \\
P22 & $C_{LbYx}$ & $1.79 \pm 0.08$ &  -  &  -  &  -  \\
P25 & $B_{L5M}$ & $1.47 \pm 0.03$ & -1 & 0.1 - 1.0 & 0.00 - 0.20 \\
T25 & $B_{MT}$ & $1.65 \pm 0.11$ & -1 & 0.1 - 5.0 & 0.00 - 0.20 \\
R26 & $B_{L5T}$ & $2.51 \pm 0.04$ & 3061 & 0.1 - 16.0 & 0.05 - 1.07 \\
R26 & $C_{L5T}$ & $1.88 \pm 0.35$ &  -  &  -  &  -  \\
R26 & $B_{L5Mg}$ & $1.23 \pm 0.01$ &  -  &  -  &  -  \\
R26 & $C_{L5Mg}$ & $2.16 \pm 0.07$ &  -  &  -  &  -  \\
R26 & $B_{MgT}$ & $2.24 \pm 0.03$ &  -  &  -  &  -  \\
R26 & $C_{MgT}$ & $-1.00 \pm 0.30$ &  -  &  -  &  -  \\
E26 & $B_{L5T}$ & $2.54 \pm 0.19$ & 44 & 0.1 - 1.0 & 0.00 - 0.05 \\
M26 & $B_{L5T}$ & $2.90 \pm 0.10$ & 155 & 0.1 - 5.5 & 0.07 - 0.20 \\
O26 & $B_{LbT}$ & $2.09 \pm 0.51$ & 27 & 3.0 - 20.0 & 0.14 - 1.17 \\
O26 & $C_{LbT}$ & $1.89 \pm 1.97$ &  -  &  -  &  -  \\
O26 & $B_{LbM}$ & $1.35 \pm 0.85$ &  -  &  -  &  -  \\
O26 & $C_{LbM}$ & $3.44 \pm 1.50$ &  -  &  -  &  -  \\
\hline \end{tabular} \end{center}
   \label{tab:pub}
   \tablefoot{ \citet[R11]{reichert11}; \citet[S20]{sereno20}; \citet[L20]{lovisari20}; \citet[L21]{lovisari21}; \citet[B22]{bahar22}; \citet[P22]{pratt22};  \citet[P25]{popesso25}; \citet[T25]{toptun25}; \citet[R26]{ramos-ceja26};\citet[E26]{eckert26}; \citet[M26]{moysan26corr}; \citet[O26]{ota26}. 
    Note that $Lb$ indicates a bolometric luminosity, $L1$ a luminosity in the 0.1--2.4 keV band, $L5$ in the 0.5--2 keV band. The other columns indicate: the number of objects in the sample ($N_{\rm obj}$; if ``-1'', the analysis is done on the stacked signal); the approximate ranges in mass ($\Delta M$) and redshift ($\Delta z$).
    }

\end{table}

\subsection{Rescaling of the thermodynamic profiles}
\label{sect:prof}

In Table~\ref{tab:norm}, I quote the exponents for the dependence on temperature (and mass) and relative redshift evolution for the thermodynamic quantities of common use.

\begin{table*}
    \caption{Dependences of the characteristic physical scales.}
\begin{center} \begin{tabular}{ccccc} \hline
$Q_{\Delta}$ & $a_T$ & $a_{T, z}$ & $a_M$ & $a_{M, z}$ \\ \hline
 $T_{\Delta}$ & $1.15 \; (0.01)$ $[1]$ & $0.11 \; (0.07)$ $[0]$ & $2/3$ \; $[2/3]$ & $2/3$ \; $[2/3]$ \\
 $n_{\Delta}$ & $0.50 \; (0.01)$ $[0]$ & $1.89 \; (0.03)$ $[2]$ & $0.29 \; (0.01)$ $[0]$ & $2.13 \; (0.04)$ $[2]$ \\
 $P_{\Delta}$ & $1.65 \; (0.01)$ $[1]$ & $2.01 \; (0.07)$ $[2]$ & $0.96 \; (0.01)$ $[2/3]$ & $2.80 \; (0.04)$ $[8/3]$ \\
 $K_{\Delta}$ & $0.82 \; (0.01)$ $[1]$ & $-1.15 \; (0.07)$ $[-4/3]$ & $0.47 \; (0.01)$ $[2/3]$ & $-0.76 \; (0.04)$ $[-2/3]$ \\
 EM$_{\Delta}$ & $1.57 \; (0.01)$ $[1/2]$ & $2.85 \; (0.07)$ $[3]$ & $0.91 \; (0.01)$ $[1/3]$ & $3.60 \; (0.09)$ $[10/3]$ \\
\hline \end{tabular} \end{center}
   \label{tab:norm}
   \tablefoot{The dependences on the temperature and redshift are written in the form $Q_{\Delta} \sim T^{a_T} E_z^{a_{Tz}}$, where ${a_T}$ and $a_{Tz}$ are functions of $f_g$ and $f_T$ as described in Tables~1 and 2 of E23. The conversion to $Q_{\Delta} \sim M^{a_M} E_z^{a_{Mz}}$ is obtained via the relation $T_{\Delta} = f_T T \sim T^{1+t_1} E_z^{t_z} = \left(E_z M\right)^{2/3}$, implying $T \sim M^{2 /(3 +3t_1)} E_z^{(2-3t_z)/(3+3t_1)}$. 
    The self-similar expectation is quoted in square brackets. }
\end{table*}

\end{appendix}
\end{document}